\documentclass[longauth, letter]{aa}  
\usepackage{graphicx}
\usepackage{wrapfig}
\usepackage{tikz}
\usepackage{lipsum}  
\usepackage{txfonts}
\usepackage{xcolor}

\bibpunct{(}{)}{;}{a}{}{,} 
\usepackage[colorlinks=true,citecolor=blue,linkcolor=magenta]{hyperref}
\usepackage{xspace}
\usepackage{float}

\newcommand{\package}[1]{\textsl{#1}}

\newcommand{\gaia}{\textsl{Gaia}\xspace}

\newcommand*{\Msun}{\ensuremath{\mathrm{M}_{\odot}\xspace}}
\newcommand*{\feh}{\ensuremath{\mathrm{[Fe/H]}\xspace}}

\newcommand{\Lz}{\ensuremath{L_{z}}\xspace}
\newcommand{\Lperp}{\ensuremath{L_{\perp}}\xspace}
\newcommand{\Etot}{\ensuremath{E_\textrm{tot}}\xspace}

\begin{document} 
   \title{\Large 
   The 33 M$_\odot$ black hole \gaia BH3 is part of the disrupted ED-2  star cluster
   }

   \author{
     E. Balbinot
     \inst{\ref{inst:kap}, \ref{inst:leiden}}
\and
     E. Dodd
     \inst{\ref{inst:kap}}
\and
     T. Matsuno
     \inst{\ref{inst:0011}}
\and
     C. Lardo\inst{\ref{inst:bologna}}
\and
     A. Helmi
     \inst{\ref{inst:kap}}
\and P.        ~Panuzzo                       \inst{\ref{inst:0001}}
\and T.        ~Mazeh                         \inst{\ref{inst:0002}}
\and B.        ~Holl                          \inst{\ref{inst:0004},\ref{inst:0005}}
\and E.        ~Caffau                        \inst{\ref{inst:0001}} 
\and A.        ~Jorissen                      \inst{\ref{inst:0007}}
\and C.        ~Babusiaux                     \inst{\ref{inst:0008}}
\and P.        ~Gavras                        \inst{\ref{inst:0009}}
\and \L{}.     ~Wyrzykowski                   \inst{\ref{inst:0012}}
\and L.        ~Eyer                          \inst{\ref{inst:0004}}
\and N.        ~Leclerc                       \inst{\ref{inst:0001}}
\and A.        ~Bombrun                       \inst{\ref{inst:0016}}
\and N.        ~Mowlavi                       \inst{\ref{inst:0004}}
\and G.M.      ~Seabroke                      \inst{\ref{inst:0018}}
\and I.        ~Cabrera-Ziri                 \inst{\ref{inst:0011}}
\and T.~M.     ~Callingham                   \inst{\ref{inst:kap}}
\and T.        ~Ruiz-Lara                    \inst{\ref{inst:0020},\ref{inst:0021}}
\and E.        ~Starkenburg                  \inst{\ref{inst:kap}}
          }

\institute{
Kapteyn Astronomical Institute, University of Groningen, Landleven 12, 9747 AD Groningen, The Netherlands\relax \label{inst:kap}
\and Leiden Observatory, Leiden University, Einsteinweg 55, 2333 CC Leiden, The Netherlands\relax \label{inst:leiden}
\and Astronomisches Rechen-Institut, Zentrum f\"{ u}r Astronomie der Universit\"{ a}t Heidelberg, M\"{ o}nchhofstr. 12-14, 69120 Heidelberg, Germany\relax                                                                                                                                                       \label{inst:0011}
\and Dipartimento di Fisica e Astronomia, Universita degli Studi di Bologna, Via Gobetti 93/2, I-40129 Bologna, Italy \label{inst:bologna}
\and GEPI, Observatoire de Paris, Universit\'{e} PSL, CNRS, 5 Place Jules Janssen, 92190 Meudon, France\relax                                                                                                                                                                                                    \label{inst:0001}
\and School of Physics and Astronomy, Tel Aviv University, Tel Aviv 6997801, Israel\relax                                                                                                                                                                                                                        \label{inst:0002}
\and Department of Astronomy, University of Geneva, Chemin Pegasi 51, 1290 Versoix, Switzerland\relax                                                                                                                                                                                                            \label{inst:0004}
\and Department of Astronomy, University of Geneva, Chemin d'Ecogia 16, 1290 Versoix, Switzerland\relax                                                                                                                                                                                                          \label{inst:0005}
\and Institut d'Astronomie et d'Astrophysique, Universit\'{e} Libre de Bruxelles CP 226, Boulevard du Triomphe, 1050 Brussels, Belgium\relax                                                                                                                                                                     \label{inst:0007}
\and Univ. Grenoble Alpes, CNRS, IPAG, 38000 Grenoble, France\relax                                                                                                                                                                                                                                              \label{inst:0008}
\and RHEA for European Space Agency (ESA), Camino bajo del Castillo, s/n, Urbanizaci\'{o}n Villafranca del Castillo, Villanueva de la Ca\~{n}ada, 28692 Madrid, Spain\relax                                                                                                                                      \label{inst:0009}
\and Astronomical Observatory, University of Warsaw,  Al. Ujazdowskie 4, 00-478 Warszawa, Poland\relax                                                                                                                                                                                                           \label{inst:0012}
\and HE Space Operations BV for European Space Agency (ESA), Camino bajo del Castillo, s/n, Urbanizaci\'{o}n Villafranca del Castillo, Villanueva de la Ca\~{n}ada, 28692 Madrid, Spain\relax                                                                                                                    \label{inst:0016}
\and Mullard Space Science Laboratory, University College London, Holmbury St Mary, Dorking, Surrey RH5 6NT, United Kingdom\relax                                                                                                                                                                                \label{inst:0018}
\and Telespazio UK S.L. for European Space Agency (ESA), Camino bajo del Castillo, s/n, Urbanizaci\'{o}n Villafranca del Castillo, Villanueva de la Ca\~{n}ada, 28692 Madrid, Spain\relax                                                                                                                        \label{inst:0019}
 \and Universidad de Granada, Departamento de Física Teórica y del Cosmos, Campus Fuente Nueva, Edificio Mecenas, E-18071,
Granada, Spain\relax
                                \label{inst:0020}
\and Instituto Carlos I de Física Te\'orica y computacional, Universidad de Granada, E-18071 Granada, Spain\relax
                                \label{inst:0021}
             }

   \date{Received \today; accepted ...}

\titlerunning{\gaia-BH3 and the ED-2 star cluster}
 
  \abstract
   {The \gaia Collaboration has recently reported the detection of a 33
   M$_\odot$ black hole in a wide binary system located in the Solar
   neighbourhood.} 
   {Here we explore the relationship between this black hole, known as
   \gaia BH3, and the nearby ED-2 halo stellar stream.}
   {We study the orbital characteristics of the \gaia BH3 binary and present
   measurements of the chemical abundances of ED-2 member stars derived from
   high-resolution spectra obtained with the VLT.}
   {We find that the Galactic orbit of the \gaia BH3 system and its metallicity
   are entirely consistent with being part of the ED-2 stream. The
   characteristics of the stream, particularly its negligible spread in
   metallicity and in other chemical elements as well as its single stellar
   population, suggest that it originated from a disrupted star cluster of low
   mass. Its age is comparable to that of the globular cluster M92 that has been
   estimated to be as old as the Universe.}
   {This is the first black hole unambiguously associated with a disrupted star
   cluster. We infer a plausible mass range for the cluster to be relatively
   narrow, between $2\times 10^3\Msun$ and $4.2\times 10^4\Msun$. This implies
   that the black hole could have formed directly from the collapse of a massive
   very-metal-poor star, but that the alternative scenario of binary
   interactions inside the cluster environment also deserves to be explored.}


   \maketitle
%

\section{Introduction}

The discovery of a 33 $\Msun$ black hole (BH) was recently reported in the Gaia
DR4 pre-release data \citep{Panuzzo:2024}. This BH is in a wide binary system
with a period of 11.6 years. Its visible companion (\gaia DR3 {\tt source\_id}
4318465066420528000) is a known high-proper motion star that is part of the
Galactic halo. The low metallicity $\feh=-2.56\pm0.12$ reported by
\citet{Panuzzo:2024} confirms the association with this Galactic component. 

This discovery is especially exciting in light of the enormous advances made
in the field of gravitational waves in recent years. Several tens of
detections of gravitational waves due to merging binary BHs have been reported
by the LIGO/VIRGO/KAGRA collaboration \citep{Abbott:2023data}. The modelling of
these events has revealed that the binary BH mass distribution follows a
power-law, with peaks at chirp masses of $\sim\!8$~\Msun\, and $\sim\!28$~\Msun
~\citep{Abbott:2023interp}. The origin of the heavier BHs is not well
understood. Because very massive stars of solar metallicity lose much of their
mass via stellar winds, it has  been argued that many of these BH could reside
in metal-poor environments such as dwarf galaxies. An alternative pathway could
be dynamical interactions in dense star clusters which may lead to hierarchical
growth of BH via BH binary mergers \citep[see e.g.][]{Antonini2020,
Fragione2023}. In this context, it is important to shed more light on the origin
of \gaia~BH3.

Since \gaia BH3 has a very retrograde and relatively loosely bound orbit,
\citet{Panuzzo:2024} have argued for a possible association with the Sequoia
accretion event \citep{Myeong:2019} identified using \gaia DR2 data. The better
astrometry available in the subsequent \gaia (E)DR3, has however revealed that
this region of integrals of motion (IoM) space, e.g. energy and angular momenta,
contains several additional substructures besides Sequoia
\citep[e.g.][]{RL2022,Dodd:2022}. Some of these substructures appear to have
distinct chemistry \citep{Matsuno2019, Naidu2020}.

Among the smaller of such retrograde structures first identified by
\citet{Dodd:2022} we highlight ED-2. This substructure has been shown to form a
dynamically cold stellar stream crossing the solar neighbourhood
\citep[][B23]{Balbinot:2023}. Because of the dynamical properties of ED-2 (a
cold but relatively wide stream) it was suggested that it could have originated
from an ultra-faint dwarf galaxy. On the other hand, the tight distribution of
its member stars in color-magnitude space and the relatively narrow (rms $\sim
0.2$~dex) metallicity distribution measured from LAMOST DR3 low resolution
spectra \citep{Li:2018VMP} for 7 stars, favoured a star cluster origin.
Interestingly the mean metallicity of ED-2 stars is
$\feh=-2.60^{+0.20}_{-0.21}$, suspiciously close to that of the companion of
\gaia BH3. 

In this Letter we demonstrate that \gaia BH3 is indeed associated with the cold
stellar stream ED-2 and that ED-2 stems from a low-mass disrupted star cluster.
Sec.~\ref{sec:kinematics} focuses on the dynamical association of \gaia BH3 with
ED-2, and Sec.~\ref{sec:spec} presents chemical abundances from follow-up
X-Shooter and UVES spectra of ED-2 members\footnote{These data had been
requested in proposals 0111.D-0263(A) (PI:Dodd) and 112.25ZW.001 (PI:Balbinot),
and hence submitted before the analyses that led to the discovery of \gaia BH3.
The co-Is of both proposals are co-authors of this paper who are not members of
the \gaia collaboration.}. In Sec.~\ref{sec:formation} we discuss the
implications of our findings and in Sec.~\ref{sec:conclusion} we present our
conclusions.

\section{Kinematics and stellar population}
\label{sec:kinematics}

\begin{figure}[h]
    \centering
    \label{fig:CMD}
    \includegraphics[width=0.85\linewidth]{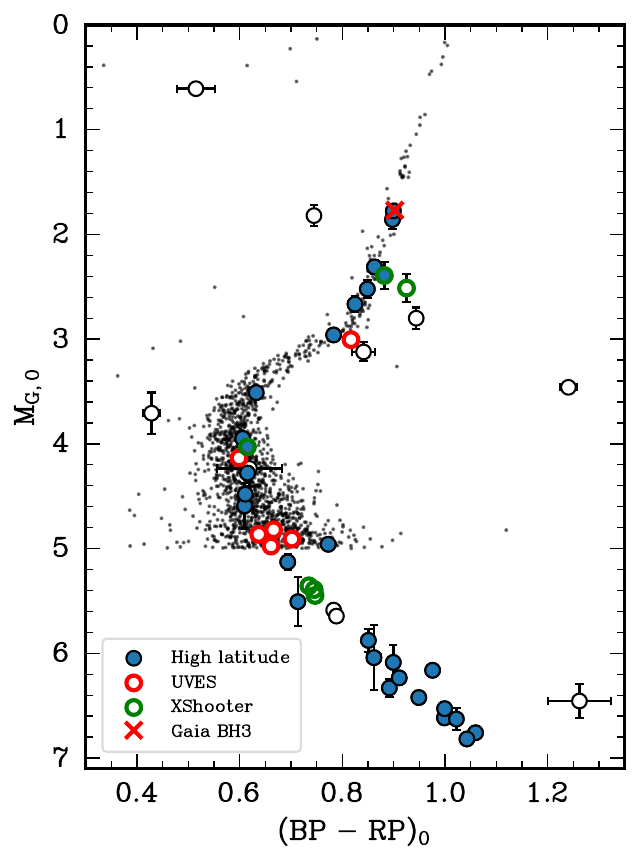}
    \caption{
    \gaia DR3 extinction corrected CMD showing the location of the
    \gaia BH3 companion (red cross) and ED-2 members (B23) as blue and empty
    circles. The former are high-latitude ($|b| > 20^{\circ}$), low-extinction
    ($E(B-V) \le 0.01$) ED-2 members. Notice the extremely tight sequence
    followed by ED-2 stars, indicative of their small metallicity dispersion.
    Their distribution is in very good agreement with the CMD of stars in the
    globular cluster M92 (truncated at $\mathrm{M}_{\mathrm{G,0}}$=5), shown in the
    background as black dots. This implies that they are of similar age, given
    their comparable metallicities. 
    The object with M$_{\mathrm{G,0}} \sim 0.6$ is an
    RR Lyrae type-c star \citep{Clementini2023}. 
}
\end{figure}

Fig.~\ref{fig:CMD} shows the extinction corrected colour-magnitude diagram (CMD)
for all ED-2 members  (see B23 for details). The cross indicates the location of
the \gaia BH3 companion star, \citep[which due to the high {\tt RUWE} value
reported in \gaia (E)DR3, was left out of the analysis by][]{Dodd:2022}. The
error bars in this figure account for the effects of distance and extinction
uncertainty. All ED-2 known member stars are within 2.5~kpc from the Sun and
their relative distance errors are smaller than 20\%. We compute the extinction
at $d \pm3~\epsilon_d$ (where $\epsilon_d$ is the distance error) to
conservatively estimate the error introduced in the 3D extinction maps of
\citet{Lallement:2022}. These uncertainties are summed in quadrature with the
photometric uncertainty. For comparison, we also plot members of the globular
cluster M92 in the background. These were selected using the method of
\citet{Vasiliev21} and are at least 4$'$ away from the M92's centre, to avoid
crowding. The CMD of M92 has been extinction corrected following the recipe
described above. 

\begin{figure}[h]
    \centering
    \includegraphics[width=0.9\linewidth]{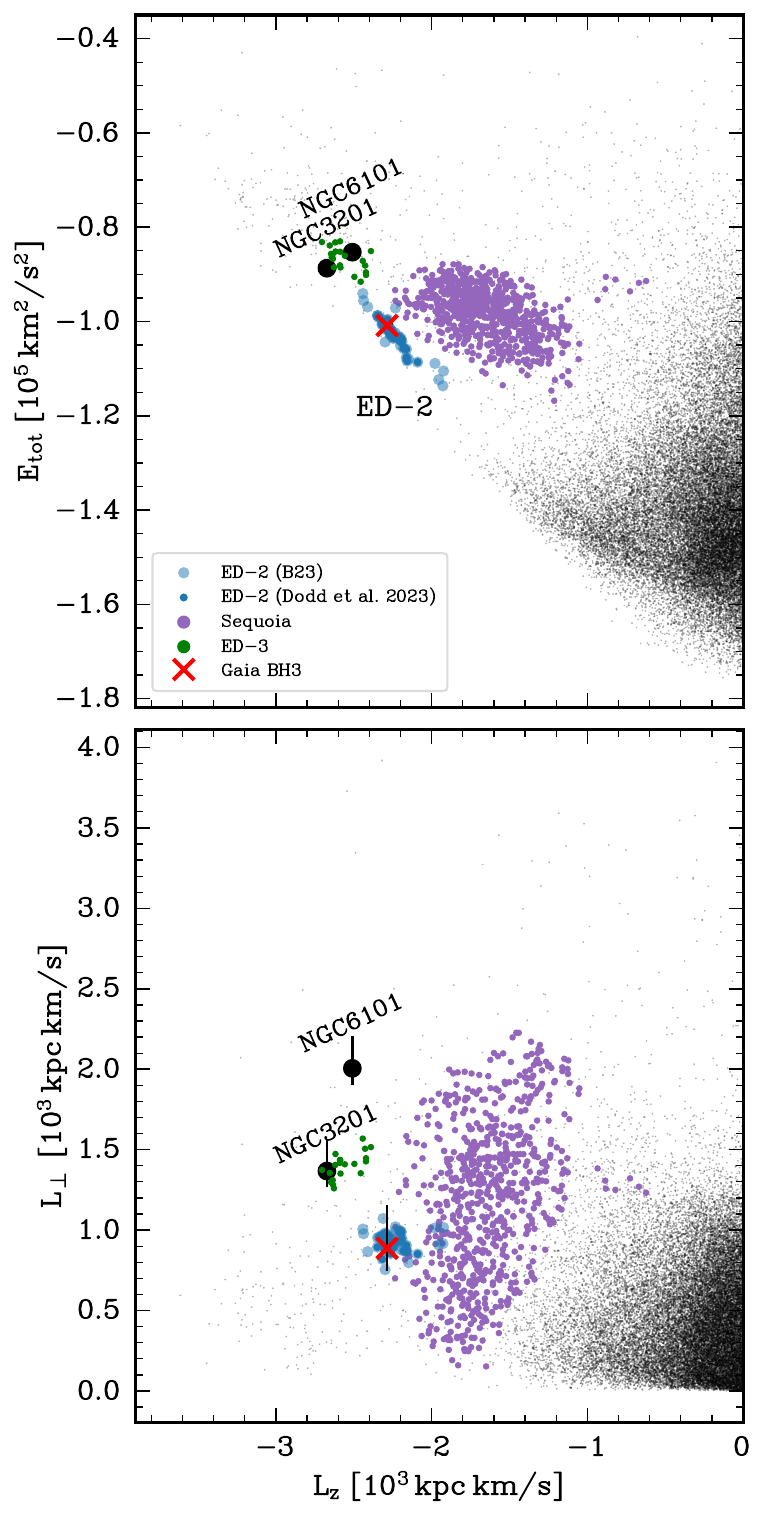}
    \caption{
     \Lz  ~{\it vs.} \Etot (top panel) and {\it vs.} \Lperp (bottom panel)
     showing ED-2 as blue and lighter blue circles, corresponding respectively
     to original members from \citet{Dodd:2022} and to the extended sample (see
     B23 for details). \gaia BH3 is shown as a red cross. Members of ED-3 and
     Sequoia \citep[as classified by][]{Dodd:2022} are also shown. The dark
     points in the background are from the \gaia DR3 6D sample within 3~kpc and
     \texttt{RUWE}~$< 1.4$. We also show two retrograde GCs. The vertical errorbars show
     the variation in $L_\perp$ for ED-2 and the two GCs along their orbits. 
     }
    \label{fig:IoM}
\end{figure}

This comparison shows that ED-2 stars match well the CMD of M92, which is known
to be one of the oldest and most metal-poor ([Fe/H]$\sim -2.3$) globular
clusters (GC) in the Galaxy \citep{Ying:2023}, with an age of 13.80 $\pm$ 0.75
Gyr. Since the main-sequence turn-off (MSTO) seems to be slightly fainter, ED-2
could potentially be even older, however, this is supported by only a single
MSTO star in ED-2. In any case, we may conclude from this comparison that ED-2
formed more than 13 Gyr ago. 

\begin{figure*}
\centering
    \includegraphics[width=0.9\linewidth]{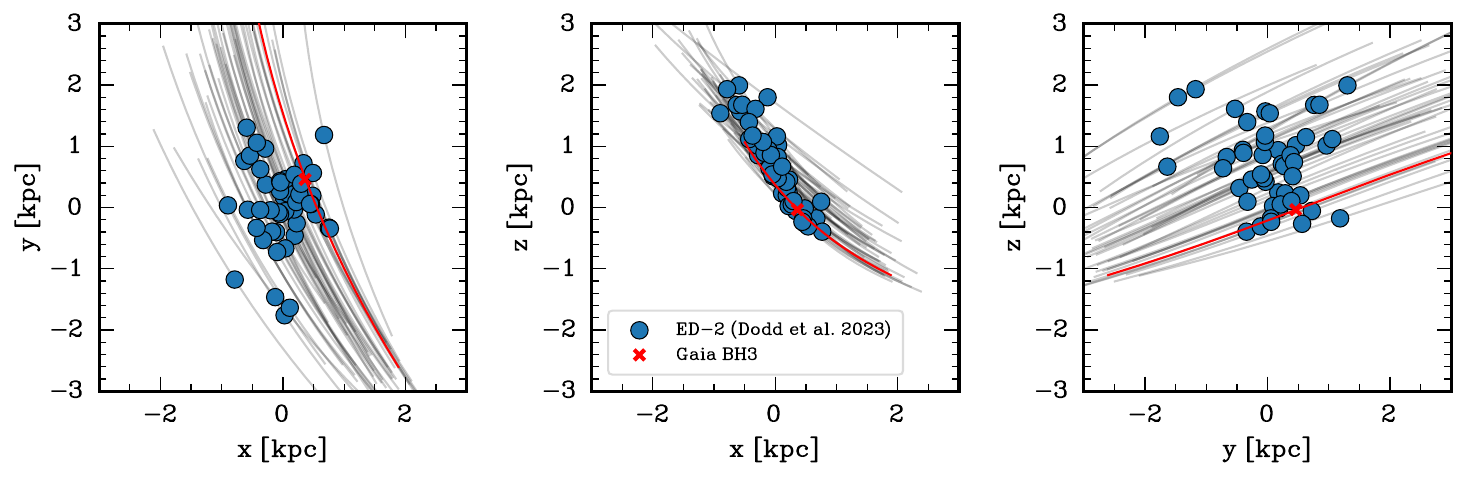}
    \caption{Cartesian heliocentric projection of the location of ED-2 members
    and their orbits  integrated in the Milky Way potential used in
    \citet{Dodd:2022} for 20 Myr. The red cross and line show the position and
    orbit of \gaia BH3, and is indistinguishable from that of the ED-2 stars. 
    }
    \label{fig:orbit}
\end{figure*}

Figure~\ref{fig:IoM} shows the distribution of ED-2 stars in IoM space:
$z$-angular momentum {\it vs.} energy (top panel), and {\it vs.} the
perpendicular component of the angular momentum (bottom panel). The location of
\gaia BH3 is indicated with a cross and falls right on top of the ED-2 stream
members. Note that the values of the IoM were computed in the Milky Way
potential of \citet{Dodd:2022}, which is slightly different from that used in
\citet{Panuzzo:2024}. The Mahalanobis distance\footnote{The Mahalanobis distance
between BH3 and ED-2 (Sequoia) is defined as $D^2_{BH3} = ({ \mu}_i-{\mu}_{BH3})^T
\Sigma_i^{-1} ({ \mu}_i-{ \mu}_{BH3})$ where ${\mu}_{BH3}$ denotes the location
of BH3 in IoM space and ${ \mu}_i$ and ${ \Sigma}_i$ are the mean and covariance
matrix of the ED-2 (Sequoia) stars.} between the centre of ED-2 and \gaia BH3 is
0.942, while that between Sequoia and \gaia BH3 is 2.718. In other words, only
17\% of the members of ED-2 are closer to its centre than \gaia BH3, while it is
in the outskirts of Sequoia as 94\% of its stars have a smaller distance. This
makes it much more likely that \gaia BH3 is associated to ED-2 than to Sequoia. 

This is further illustrated by the trajectories followed by the stars in ED-2
and \gaia BH3 shown in Fig.~\ref{fig:orbit}, where there is no noticeable
distinction between the different objects. It is interesting that BH3 is not at
the centre of the distribution of stars. Whether this is real or due to
incompleteness in the sample (i.e. the distance limit and the magnitude limit of
the RVS dataset) should be scrutinised in-depth in further studies. 

Given the size of the ED-2 structure in IoM space, which is rather comparable to
that of other globular clusters, such as NCGC3201 and NGC6101
also shown in Fig.~\ref{fig:IoM}, we tentatively conclude that ED-2 stems from a
GC-like progenitor. The good fit obtained from a single stellar population
further supports this conclusion. 
   \begin{figure}[h!]
   \centering
   \includegraphics[width=\linewidth]{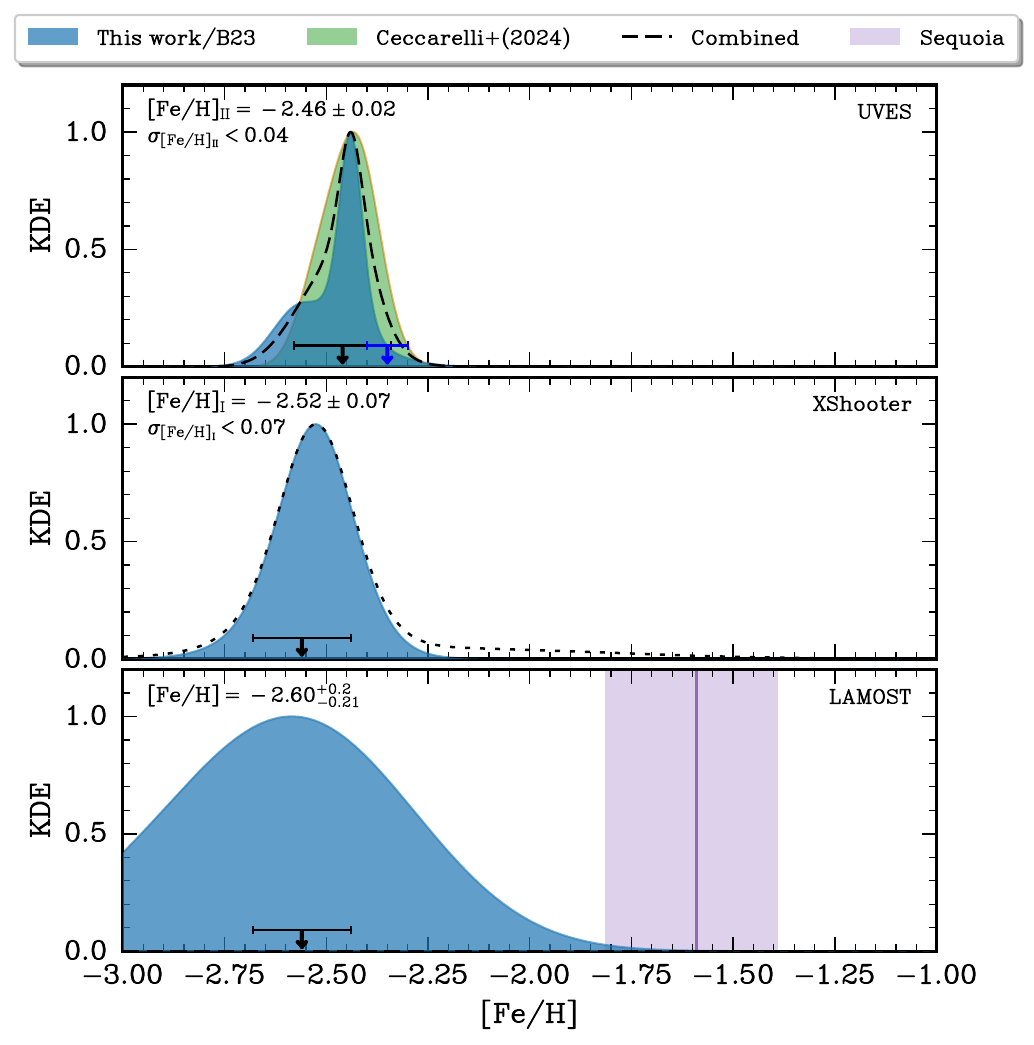}
   \caption{Metallicity distribution for ED-2 based on UVES, X-Shooter and
   LAMOST spectra in the top, middle and bottom panels respectively. The arrow
   and errorbar show \gaia BH3's companion's metallicity and uncertainty, as
   inferred by \citet{Panuzzo:2024}, and in the top panel, using \ion{Fe}{II}
   lines and by our own analysis in blue (see also Table~\ref{table:1}). In the
   top left of each panel the best-fit metallicity and the upper bound on the
   metallicity dispersion are given. In the middle panel, the dashed
   distribution includes a star with large [Fe/H] uncertainties. We also show in
   the bottom panel the 25\%--75\% quantiles for members of Sequoia
   \citep[following the classification of][]{Dodd:2022} as a shaded band using
   metallicity estimates
   from LAMOST. 
}
   \label{fig:spectra}%
   \end{figure}

\section{Chemical abundances}
  \label{sec:spec}

We obtained spectra for 3 stars as part of the follow-up of the ED-2 stream in
period 111 (April -- September 2023; proposal submitted in September 2022) with
the optical spectrograph, UVES \citep{Dekker:2000} mounted at the Very Large
Telescope (VLT) of the European Southern Observatory (ESO). We have also used
ESO archival data for another ED-2 member, {\tt source\_ID} 4479226310758314496.
Additionally, we observed 6 ED-2 core members with X-Shooter \citep{Vernet:2011}
at the VLT in period 112 (October 2023 -- March 2024; proposal submitted in
March 2023). In all cases, we used the phase-3 data products provided by ESO for
further analysis. In the Appendix we provide details of the observational
set-ups, and we also describe the procedure used to derive the stellar
parameters and chemical abundances of the UVES stars and the [Fe/H] for the
X-Shooter targets. We list the results in Table~\ref{table:1}  and
Table~\ref{tab:specXS} for the stars observed with UVES and X-shooter
respectively.

Fig.~\ref{fig:spectra} shows the metallicity distribution derived for the ED-2
stars in our programs. The top panel corresponds to the UVES targets whose
metallicity is measured from the \ion{Fe}{II} lines, which are more reliable due
to their small sensitivity to the adopted stellar parameters and non-LTE
effects. The middle panels are for the X-Shooter stars, while the bottom panel
shows the distribution derived by B23 compared to that of Sequoia as defined by
\citet{Dodd:2022}, both using LAMOST spectra. The black arrow and errorbar show
\gaia BH3 visible companion’s metallicity and its uncertainty. This figure
confirms, now on the basis of the metallicity, that the black hole is a member
of ED-2, and has a negligible probability to be part of Sequoia.

We measure the mean metallicity and metallicity dispersion ($\sigma_{\feh,
int}$) of ED-2 in the UVES and X-Shooter samples assuming a simple normal
distribution with dispersion $\sigma^2$ = $\smash{\sigma_{\feh, int}^2}$ +
$\smash{\sigma^2_{\feh , j}}$ , i.e. the sum in quadrature of an intrinsic
dispersion and the metallicity uncertainty in each $j$-th data point. We use
this distribution to maximize the likelihood of our model using \textsc{emcee}
\citep{emcee}. For the UVES sample, we find a best-fit metallicity of $\feh_{\rm
II}=-2.46\pm0.02$ and a metallicity dispersion $\sigma_{\feh} < 0.04$.
Similarly, for the X-Shooter sample, we find a best-fit metallicity of
$\feh=-2.51\pm0.07$ and a metallicity dispersion $\sigma_{\feh} <
0.07$\footnote{Although the star with {\tt source\_ID} 3757312745743087232 is
more metal-rich than the remainder of the sample (see Table~\ref{tab:specXS}),
it has very large uncertainties, and its inclusion has no effect in the derived
mean and spread in metallicity.}. The uncertainties in metallicity were computed
from the standard deviation of the posterior distribution, while the upper
limits in $\sigma_{\feh}$ are the 67\% quantile of the posterior.  We thus find
that the intrinsic metallicity dispersion of ED-2 is consistent with zero. This
favours a star cluster origin as opposed to a dwarf galaxy, as even ultra-faint
dwarfs have a scatter of at least 0.3 dex \citep{Simon:2019}.

\begin{figure}
    \includegraphics[width=0.9\linewidth]{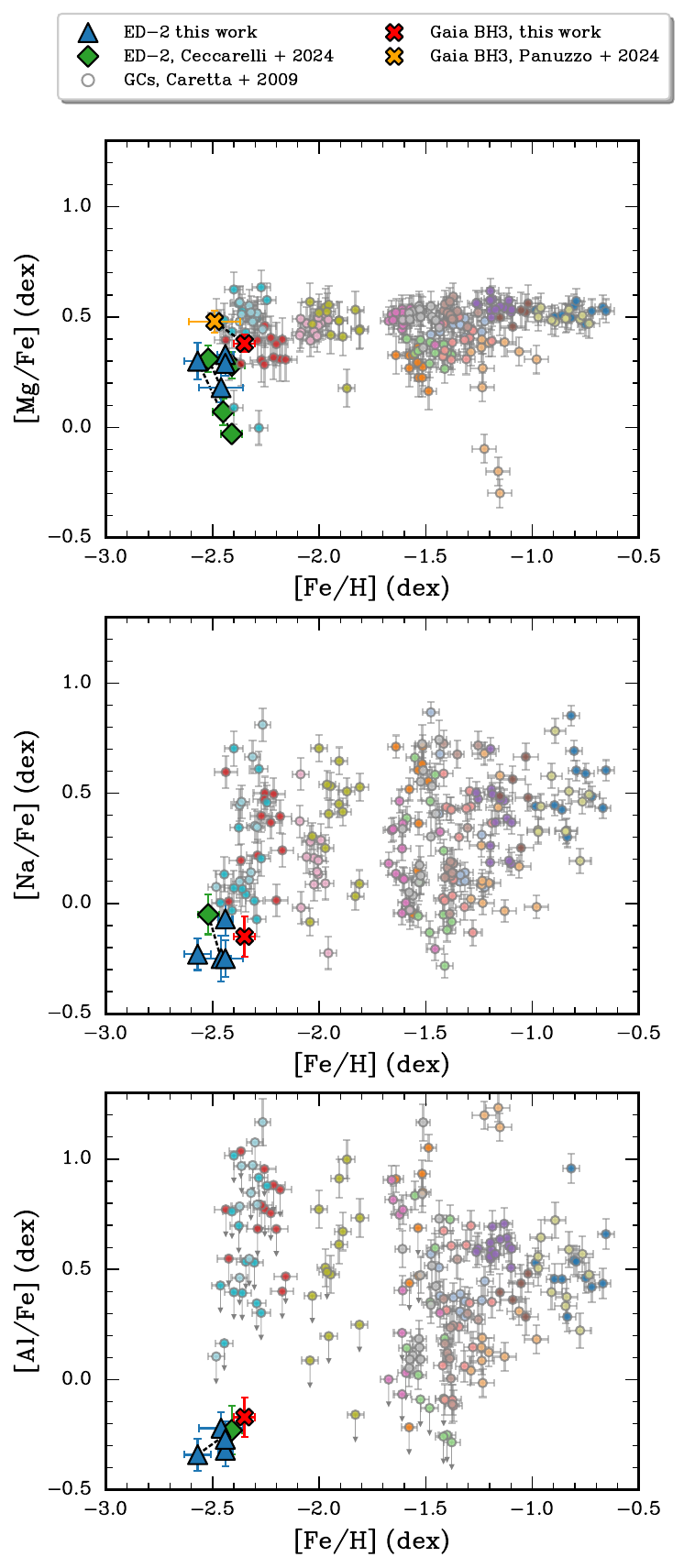}
    \caption{Abundances of Mg/Fe, Na/Fe and Al/Fe for the 4 ED-2 stars in our
    UVES sample (blue triangles), and for ED-2 stars from \citet[][green
    diamonds]{Ceccarelli2024}. Note the good agreement and small scatter. Also
    chemically the companion of \gaia BH3 (cross symbol, red corresponds to our
    own abundances and orange to the measurements from \citet{Panuzzo:2024}, see
    also Table \ref{table:1}) is indistinguishable from ED-2 stars (see also
    text). For comparison, we have plotted also (upper limits to) the abundances
    for several GCs from \citet{carretta2009UVES}.}
    \label{fig:abund}
\end{figure}

Fig.~\ref{fig:abund} shows the abundances of Mg, Na and Al with respect to Fe
for the stars observed with UVES (blue triangles). The scatter in all elements
is very small, again indicating that the ED-2 originated in a star cluster. Also
the abundances measured for other ED-2 stars by \citet[][green
diamonds]{Ceccarelli2024} show very comparable values. The measurements for the
companion star of \gaia BH3 as provided by \citet[][in orange]{Panuzzo:2024} and
by our own analysis (in red) are shown with a cross symbol, and are fully
consistent with those of the ED-2 stars. Also its measured [Eu/Fe]=0.52 is in
excellent agreement with that of another star in ED-2, for which we could
measure [Eu/Fe] = 0.61, a value that supports similar amounts of r-process
enhancement across the system. ED-2's mean abundance of [Ba/Fe]~$\sim -0.22$
(and its small dispersion of $0.1$ dex) is consistent with that of other halo
stars, but different from that seen in ultra-faint dwarf galaxies, which
typically depict much larger or much lower values \citep{Ji2019}. In
Fig.~\ref{fig:abund} we plotted for comparison the abundances of a set of GCs
from \citet[][all of which are more metal-rich]{carretta2009UVES}, which reveal
a similar scatter in [Mg/Fe] as ED-2 members but larger in Na and Al.

The low [Al/Fe] and high [Mg/Mn] of the ED-2 stars and of the companion star of
\gaia BH3 (see Table~\ref{table:1}) places ED-2 members in a region of
abundance-space that is referred to as ``chemically unevolved''
\citep{Hawkins2015, Fernandes2023}. This could hint at an accretion origin of
ED-2 given also its highly retrograde orbit. However, care must be taken when
interpreting this chemical space since its validity as an indicator of a possible
accretion origin has not been firmly established for star clusters.

\section{Discussion}
  \label{sec:formation}

Having established that \gaia BH3 formed in a star cluster, we now explore
possible formation channels. We also attempt to infer some properties of the
ED-2 parent cluster. Note that these findings naturally explain the ``normal''
chemical composition of its accompanying star, in the sense that the binary
could easily have formed in the cluster after the BH was born.

The most straightforward formation scenario for BHs is through the collapse of a
very massive star. The mass of such a BH is dictated by the star's mass at the
end of its evolution. Due to the details of the mass-loss process, this can
differ significantly from its initial value. Using the single-star initial-final
mass relations (IFMR) from \citet{Fryer:2012} implemented in {\tt ssptools}
\footnote{\url{https://github.com/SMU-clusters/ssptools}}
\citep{Balbinot:2018,Dickson:2023} and a Kroupa initial mass function, we can
infer how many stellar BHs of a given mass are likely to form as a function of the
cluster mass. We find that the minimum mass for a star cluster to host at least
1 stellar BH of the size of \gaia BH3 or higher is $M_{\rm cl,min} \sim
2\times10^3$~\Msun. In this case, \gaia BH3 would be a first generation BH.

Alternative pathways to produce very massive BHs have been proposed that require
binary evolution and dynamical hardening of these binary systems. These
processes take place in dense stellar systems such as GCs \citep[see
e.g.][]{SPZ:2000}, and see also the recent work on young star clusters by
\citet{Rastello:2023, Tanikawa2024, DiCarlo2024}. Due to their stochastic
nature, such processes can produce BH with a wide range of masses \citep[see
e.g.][]{Antonini2020}. In this case, \gaia BH3 could have formed via the mergers
of subsequent generations of BHs.

The scatter in the Na and Al abundances seen in GCs is indicative of multiple
stellar populations and has a dependence on both mass and metallicity \citep[see
e.g.][]{Gratton2019}. Fnx I, a GC in the Fornax dwarf spheroidal, has
[Fe/H]$\sim -2.5$, and does show multiple populations, as well as scatter in
[Na/Fe] \citep{Letarte2006}, and its initial mass has been estimated to be
4.2$\times10^4\Msun$ by \citet{Fnx1mass2016}. Therefore, the (near) lack of such
a scatter for ED-2 suggests its parent cluster was lighter in mass than Fnx~I.
We may also use the relationship by \citet{Pancino2017} between the spread in
[Al/Fe], the mean [Fe/H] of the cluster and its mass: $\Delta[{\rm Al/Mg}] =
0.67(\pm 0.21)\log M_{\rm cl}- 0.53( \pm 0.17) [{\rm Fe/H}]- 3.16(\pm1.11).$ By
randomly drawing [Al/Mg] for each star within the uncertainties, we obtain a
distribution of $\Delta$[Al/Mg], and considering the uncertainties in the
coefficients in a similar fashion, we can infer a distribution of possible
cluster masses. We find a median value of $\log M_{\rm cl}/\Msun = 2.86$, with
the 25\% and 75\% quantiles of $\log M_{\rm cl}/\Msun$ being 1.68 and 4.33
respectively. If the scatter is only due to errors, this estimate would be an
upper limit. It is however based on an extrapolation of a relation determined
for Galactic clusters whose metallicities are all higher than ED-2's, and whose
present-day masses could differ from their initial values by at least a factor
of 2 \citep{Anders:2009}. Nevertheless, it is reassuring that this upper limit
is consistent with that provided by the comparison to Fnx~I. 

Further support for ED-2 having been a low-mass star cluster comes from the fact
that the \gaia BH3 system is a relatively wide binary, with a period of 11.6 yr.
Such long period binaries do not survive in massive GCs because they are either
quickly disrupted or become tighter because of interactions within such systems
\citep[][although its low mass-ratio enhances the chances of
survival]{Ivanova:2005}. Unlike NGC6101 or NGC3201, the two GC with similar
orbits plotted in Fig.~\ref{fig:IoM}, ED-2 did not manage to survive as a star
cluster until the present day. This could be due to its lower mass or a lower
density. But also the retention of \gaia BH3 by the ED-2 cluster could have
contributed in speeding-up its disruption
\citep{Gieles:2021}.

\section{Conclusions}
  \label{sec:conclusion}

In this Letter, we have shown that the 33~$\Msun$ black hole \gaia BH3 is
associated to the ED-2 retrograde halo stellar stream. The BH's orbit around the
Galaxy is indistinguishable from that of ED-2 members. Using high-resolution
spectra of ED-2 stars, we have determined that ED-2's mean metallicity is
entirely consistent with that of the companion of \gaia BH3, as are other
chemical elemental abundances such as [Mg/Fe],  [Eu/Fe] and [Ba/Fe].
Furthermore, we have shown that the metallicity spread in ED-2 is consistent
with zero, indicating that it stems from a disrupted star cluster. This is
entirely in-line with its colour-magnitude diagram, which is very well fit by an
extremely old single stellar population, similar to that of the GC M92,
indicating that the progenitor of \gaia BH3 formed more than 13~Gyr ago. The
(near) lack of scatter in Na and Al suggests that ED-2's parent system was a
small cluster with mass smaller than $ 4.2 \times 10^4 \Msun$. This would leave
a small window for \gaia BH3 to be the direct result of the collapse of a
massive star, since we have found that such a heavy BH can only form in a system
more massive than $2\times 10^3\Msun$. To shed more light on its formation
channels, sophisticated dynamical models of the ED-2 parent cluster, including
stellar evolution and binary interactions, and using as boundary conditions
those inferred in this paper (e.g.~mass range, metallicity, and orbit) are
needed. Furthermore, the mapping of the ED-2 stream beyond the Solar
neighbourhood would allow a reliable and independent determination of the
initial cluster mass. Finally, detailed chemical abundances for more of its
members would put a tighter constraint on the lack of a spread of light elements
and constrain further the evolution of the system.

\begin{acknowledgements}
The authors would like to thank Elena Pancino and Mark Gieles for the insightful
discussion on GC chemistry and BH evolution, and Gijs Nelemans for useful
references. We acknowledge support from a Spinoza prize from the Netherlands
Organisation for Scientific Research (NWO). TM is supported by a Gliese
Fellowship at the Zentrum f\"{u}r Astronomie, University of Heidelberg, Germany.
This study was supported by the Klaus Tschira Foundation. We have made use of
data from the European Space Agency (ESA) mission \gaia
(https://www.cosmos.esa.int/gaia), processed by the \gaia Data Processing and
Analysis Consortium (DPAC, https://www.cosmos.esa.int/web/gaia/dpac/consortium).
Funding for the DPAC has been provided by national institutions, in particular
the institutions participating in the \gaia Multilateral Agreement. Non-public
data underlying this article will be shared on reasonable request to the
authors. Based on observations made with ESO Telescopes at the La Silla Paranal
Observatory under programme ID 112.25ZW.001 (PI: Balbinot), 111.2537.001 (PI:
Dodd), and 106.21JJ.001 (PI: Matsuno).

The following software packages where used in this publication:
\package{Astropy} \citep{astropy, astropy:2018},
\package{dustmaps} \citep{Green:2018},
\package{IPython} \citep{ipython},
\package{matplotlib} \citep{mpl},
\package{numpy} \citep{numpy},
\package{scipy} \citep{scipy},
\package{vaex} \citep{vaex}
\end{acknowledgements}


\bibliographystyle{aa} 
\bibliography{refs.bib} 

\begin{thebibliography}{59}
\expandafter\ifx\csname natexlab\endcsname\relax\def\natexlab#1{#1}\fi

\bibitem[{{Abbott} {et~al.}(2023{\natexlab{a}}){Abbott}, {Abbott}, {Acernese},
  {Ackley}, {Adams}, {Adhikari}, {Adhikari}, {Adya}, {Affeldt}, {Agarwal}, \&
  et~al.}]{Abbott:2023data}
{Abbott}, R., {Abbott}, T.~D., {Acernese}, F., {et~al.} 2023{\natexlab{a}},
  Physical Review X, 13, 041039

\bibitem[{{Abbott} {et~al.}(2023{\natexlab{b}}){Abbott}, {Abbott}, {Acernese},
  {Ackley}, {Adams}, {Adhikari}, {Adhikari}, {Adya}, {Affeldt}, {Agarwal}, \&
  et~al.}]{Abbott:2023interp}
{Abbott}, R., {Abbott}, T.~D., {Acernese}, F., {et~al.} 2023{\natexlab{b}},
  Physical Review X, 13, 011048

\bibitem[{{Anders} {et~al.}(2009){Anders}, {Lamers}, \&
  {Baumgardt}}]{Anders:2009}
{Anders}, P., {Lamers}, H.~J.~G.~L.~M., \& {Baumgardt}, H. 2009, \aap, 502, 817

\bibitem[{{Andrae} {et~al.}(2023){Andrae}, {Rix}, \& {Chandra}}]{Andrae:2023}
{Andrae}, R., {Rix}, H.-W., \& {Chandra}, V. 2023, arXiv e-prints,
  arXiv:2302.02611

\bibitem[{{Antonini} \& {Gieles}(2020)}]{Antonini2020}
{Antonini}, F. \& {Gieles}, M. 2020, \mnras, 492, 2936

\bibitem[{{Astropy Collaboration} {et~al.}(2018){Astropy Collaboration},
  {Price-Whelan}, {Sip{ \H{o}}cz }, {G{\"u}nther}, {Lim}, {Crawford},
  {Conseil}, {Shupe}, {Craig}, { Dencheva}, {Ginsburg}, {VanderPlas}, {Bradley
  }, {P{\'e}rez-Su{ \'a}rez}, {de Val-Borro}, { Aldcroft}, {Cruz},
  {Robitaille}, { Tollerud}, {Ardelean}, {Babej}, {Bach}, {Bachetti},
  {Bakanov}, {Bamford}, { Barentsen}, {Barmby}, {Baumbach}, { Berry},
  {Biscani}, {Boquien}, { Bostroem}, { Bouma}, {Brammer}, {Bray}, {
  Breytenbach}, {Buddelmeijer}, {Burke}, { Calderone}, { Cano Rodr{\' \i}
  guez}, {Cara}, {Cardoso}, { Cheedella}, {Copin}, { Corrales}, { Crichton},
  {D'Avella}, {Deil}, {Depagne }, { Dietrich}, {Donath}, { Droettboom}, {Earl},
  {Erben}, {Fabbro}, { Ferreira}, {Finethy}, {Fox}, {Garrison}, {Gibbons},
  {Goldstein}, {Gommers}, {Greco}, { Greenfield }, {Groener}, {Grollier},
  {Hagen}, {Hirst}, {Homeier}, {Horton}, { Hosseinzadeh}, {Hu}, {Hunkeler}, {
  Ivezi{\'c}}, {Jain}, {Jenness}, { Kanarek}, { Kendrew}, {Kern}, {
  Kerzendorf}, {Khvalko}, {King}, { Kirkby}, {Kulkarni}, {Kumar}, {Lee},
  {Lenz}, {Littlefair}, {Ma}, { Macleod}, { Mastropietro}, {McCully}, {
  Montagnac}, {Morris}, {Mueller}, { Mumford}, {Muna}, { Murphy}, { Nelson},
  {Nguyen }, {Ninan}, {N{ \" o}the}, {Ogaz}, { Oh}, {Parejko}, {Parley},
  {Pascual}, { Patil}, { Patil }, {Plunkett}, {Prochaska}, {Rastogi}, {Reddy
  Janga}, {Sabater}, { Sakurikar}, { Seifert}, {Sherbert}, { Sherwood-Taylor},
  { Shih}, {Sick}, { Silbiger}, {Singanamalla}, {Singer}, {Sladen}, {Sooley},
  {Sornarajah}, {Streicher}, { Teuben}, {Thomas}, {Tremblay}, {Turner},
  {Terr{\'o}n}, {van Kerkwijk}, {de la Vega}, { Watkins}, {Weaver}, {Whitmore},
  {Woillez }, {Zabalza}, \& {Astropy Contributors}}]{astropy:2018}
{Astropy Collaboration}, {Price-Whelan}, A.~M., {Sip{ \H{o}}cz }, B.~M.,
  {et~al.} 2018, \aj, 156, 123

\bibitem[{{Astropy Collaboration} {et~al.}(2013){Astropy Collaboration},
  {Robitaille}, { Tollerud}, {Greenfield}, {Droettboom}, {Bray}, {Aldcroft},
  {Davis}, {Ginsburg}, { Price-Whelan}, {Kerzendorf}, {Conley}, { Crighton},
  {Barbary}, {Muna}, { Ferguson}, {Grollier}, { Parikh}, {Nair}, {Unther},
  {Deil}, {Woillez}, { Conseil}, { Kramer}, {Turner}, {Singer}, {Fox},
  {Weaver}, {Zabalza}, {Edwards}, { Azalee Bostroem}, {Burke}, {Casey},
  {Crawford}, {Dencheva}, {Ely}, {Jenness}, {Labrie}, {Lim}, {Pierfederici},
  {Pontzen}, {Ptak}, {Refsdal}, {Servillat}, \& {Streicher}}]{astropy}
{Astropy Collaboration}, {Robitaille}, T.~P., { Tollerud}, E.~J., {et~al.}
  2013, \aap, 558, A33

\bibitem[{{Balbinot} \& {Gieles}(2018)}]{Balbinot:2018}
{Balbinot}, E. \& {Gieles}, M. 2018, \mnras, 474, 2479

\bibitem[{{Balbinot} {et~al.}(2023){Balbinot}, {Helmi}, {Callingham},
  {Matsuno}, {Dodd}, \& {Ruiz-Lara}}]{Balbinot:2023}
{Balbinot}, E., {Helmi}, A., {Callingham}, T., {et~al.} 2023, \aap, 678, A115

\bibitem[{{Breddels} \& {Veljanoski}(2018)}]{vaex}
{Breddels}, M.~A. \& {Veljanoski}, J. 2018, \aap, 618, A13

\bibitem[{{Carretta} {et~al.}(2009){Carretta}, {Bragaglia}, {Gratton}, \&
  {Lucatello}}]{carretta2009UVES}
{Carretta}, E., {Bragaglia}, A., {Gratton}, R., \& {Lucatello}, S. 2009, \aap,
  505, 139

\bibitem[{{Casagrande} \& {VandenBerg}(2014)}]{Casagrande2014}
{Casagrande}, L. \& {VandenBerg}, D.~A. 2014, \mnras, 444, 392

\bibitem[{{Ceccarelli} {et~al.}(2024){Ceccarelli}, {Massari}, {Mucciarelli},
  {Bellazzini}, {Nunnari}, {Cusano}, {Lardo}, {Romano}, {Ilyin}, \&
  {Stokholm}}]{Ceccarelli2024}
{Ceccarelli}, E., {Massari}, D., {Mucciarelli}, A., {et~al.} 2024, \aap, 684,
  A37

\bibitem[{{Clementini} {et~al.}(2023){Clementini}, {Ripepi}, {Garofalo},
  {Molinaro}, {Muraveva}, {Leccia}, {Rimoldini}, {Holl}, {Jevardat de
  Fombelle}, {Sartoretti}, {Marchal}, {Audard}, {Nienartowicz}, {Andrae},
  {Marconi}, {Szabados}, {Evans}, {Lecoeur-Taibi}, {Mowlavi}, {Musella}, \&
  {Eyer}}]{Clementini2023}
{Clementini}, G., {Ripepi}, V., {Garofalo}, A., {et~al.} 2023, \aap, 674, A18

\bibitem[{{de Boer} \& {Fraser}(2016)}]{Fnx1mass2016}
{de Boer}, T.~J.~L. \& {Fraser}, M. 2016, \aap, 590, A35

\bibitem[{{Dekker} {et~al.}(2000){Dekker}, {D'Odorico}, {Kaufer}, {Delabre}, \&
  {Kotzlowski}}]{Dekker:2000}
{Dekker}, H., {D'Odorico}, S., {Kaufer}, A., {Delabre}, B., \& {Kotzlowski}, H.
  2000, in Society of Photo-Optical Instrumentation Engineers (SPIE) Conference
  Series, Vol. 4008, Optical and IR Telescope Instrumentation and Detectors,
  ed. M.~{Iye} \& A.~F. {Moorwood}, 534--545

\bibitem[{{Di Carlo} {et~al.}(2024){Di Carlo}, {Agrawal}, {Rodriguez}, \&
  {Breivik}}]{DiCarlo2024}
{Di Carlo}, U.~N., {Agrawal}, P., {Rodriguez}, C.~L., \& {Breivik}, K. 2024,
  \apj, 965, 22

\bibitem[{{Dickson} {et~al.}(2023){Dickson}, {H{\'e}nault-Brunet}, {Baumgardt},
  {Gieles}, \& {Smith}}]{Dickson:2023}
{Dickson}, N., {H{\'e}nault-Brunet}, V., {Baumgardt}, H., {Gieles}, M., \&
  {Smith}, P.~J. 2023, \mnras, 522, 5320

\bibitem[{{Dodd} {et~al.}(2023){Dodd}, {Callingham}, {Helmi}, {Matsuno},
  {Ruiz-Lara}, {Balbinot}, \& {L{\"o}vdal}}]{Dodd:2022}
{Dodd}, E., {Callingham}, T.~M., {Helmi}, A., {et~al.} 2023, \aap, 670, L2

\bibitem[{{Fernandes} {et~al.}(2023){Fernandes}, {Mason}, {Horta}, {Schiavon},
  {Hayes}, {Hasselquist}, {Feuillet}, {Beaton}, {J{\"o}nsson}, {Kisku},
  {Lacerna}, {Lian}, {Minniti}, \& {Villanova}}]{Fernandes2023}
{Fernandes}, L., {Mason}, A.~C., {Horta}, D., {et~al.} 2023, \mnras, 519, 3611

\bibitem[{{Foreman-Mackey} {et~al.}(2013){Foreman-Mackey}, {Hogg}, {Lang}, \&
  {Goodman}}]{emcee}
{Foreman-Mackey}, D., {Hogg}, D.~W., {Lang}, D., \& {Goodman}, J. 2013, \pasp,
  125, 306

\bibitem[{{Fragione} \& {Rasio}(2023)}]{Fragione2023}
{Fragione}, G. \& {Rasio}, F.~A. 2023, \apj, 951, 129

\bibitem[{{Fryer} {et~al.}(2012){Fryer}, {Belczynski}, {Wiktorowicz},
  {Dominik}, {Kalogera}, \& {Holz}}]{Fryer:2012}
{Fryer}, C.~L., {Belczynski}, K., {Wiktorowicz}, G., {et~al.} 2012, \apj, 749,
  91

\bibitem[{{Gaia Collaboration: Panuzzo} {et~al.}(2024){Gaia Collaboration:
  Panuzzo}, {Mazeh, T.}, \& {et al., }}]{Panuzzo:2024}
{Gaia Collaboration: Panuzzo}, {Mazeh, T.}, \& {et al., }. 2024, A\&A

\bibitem[{{Gieles} {et~al.}(2021){Gieles}, {Erkal}, {Antonini}, {Balbinot}, \&
  {Pe{\~n}arrubia}}]{Gieles:2021}
{Gieles}, M., {Erkal}, D., {Antonini}, F., {Balbinot}, E., \& {Pe{\~n}arrubia},
  J. 2021, Nature Astronomy, 5, 957

\bibitem[{{Gratton} {et~al.}(2019){Gratton}, {Bragaglia}, {Carretta},
  {D'Orazi}, {Lucatello}, \& {Sollima}}]{Gratton2019}
{Gratton}, R., {Bragaglia}, A., {Carretta}, E., {et~al.} 2019, \aapr, 27, 8

\bibitem[{{Green}(2018)}]{Green:2018}
{Green}, G. 2018, The Journal of Open Source Software, 3, 695

\bibitem[{{Green} {et~al.}(2019){Green}, {Schlafly}, {Zucker}, {Speagle}, \&
  {Finkbeiner}}]{Green2019}
{Green}, G.~M., {Schlafly}, E., {Zucker}, C., {Speagle}, J.~S., \&
  {Finkbeiner}, D. 2019, \apj, 887, 93

\bibitem[{{Gustafsson} {et~al.}(2008){Gustafsson}, {Edvardsson}, {Eriksson},
  {J{\o}rgensen}, {Nordlund}, \& {Plez}}]{Gustafsson2008}
{Gustafsson}, B., {Edvardsson}, B., {Eriksson}, K., {et~al.} 2008, \aap, 486,
  951

\bibitem[{{Hawkins} {et~al.}(2015){Hawkins}, {Jofr{\'e}}, {Masseron}, \&
  {Gilmore}}]{Hawkins2015}
{Hawkins}, K., {Jofr{\'e}}, P., {Masseron}, T., \& {Gilmore}, G. 2015, \mnras,
  453, 758

\bibitem[{{Hunter}(2007)}]{mpl}
{Hunter}, J.~D. 2007, Computing in Science and Engineering, 9, 90

\bibitem[{{Ivanova} {et~al.}(2005){Ivanova}, {Belczynski}, {Fregeau}, \&
  {Rasio}}]{Ivanova:2005}
{Ivanova}, N., {Belczynski}, K., {Fregeau}, J.~M., \& {Rasio}, F.~A. 2005,
  \mnras, 358, 572

\bibitem[{{Ji} {et~al.}(2019){Ji}, {Simon}, {Frebel}, {Venn}, \&
  {Hansen}}]{Ji2019}
{Ji}, A.~P., {Simon}, J.~D., {Frebel}, A., {Venn}, K.~A., \& {Hansen}, T.~T.
  2019, \apj, 870, 83

\bibitem[{Jones {et~al.}(2001--)Jones, Oliphant, Peterson, {et~al.}}]{scipy}
Jones, E., Oliphant, T., Peterson, P., {et~al.} 2001--, {SciPy}: Open source
  scientific tools for {Python}

\bibitem[{{Kupka} {et~al.}(2011){Kupka}, {Dubernet}, \& {VAMDC
  Collaboration}}]{vald}
{Kupka}, F., {Dubernet}, M.~L., \& {VAMDC Collaboration}. 2011, Baltic
  Astronomy, 20, 503

\bibitem[{{Kurucz}(2005)}]{kurucz05}
{Kurucz}, R.~L. 2005, Memorie della Societa Astronomica Italiana Supplementi,
  8, 14

\bibitem[{{Lallement} {et~al.}(2022){Lallement}, {Vergely}, {Babusiaux}, \&
  {Cox}}]{Lallement:2022}
{Lallement}, R., {Vergely}, J.~L., {Babusiaux}, C., \& {Cox}, N.~L.~J. 2022,
  \aap, 661, A147

\bibitem[{{Letarte} {et~al.}(2006){Letarte}, {Hill}, {Jablonka}, {Tolstoy},
  {Fran{\c{c}}ois}, \& {Meylan}}]{Letarte2006}
{Letarte}, B., {Hill}, V., {Jablonka}, P., {et~al.} 2006, \aap, 453, 547

\bibitem[{{Li} {et~al.}(2018){Li}, {Tan}, \& {Zhao}}]{Li:2018VMP}
{Li}, H., {Tan}, K., \& {Zhao}, G. 2018, \apjs, 238, 16

\bibitem[{{Lind} {et~al.}(2022){Lind}, {Nordlander}, {Wehrhahn}, {Montelius},
  {Osorio}, {Barklem}, {Af{\c{s}}ar}, {Sneden}, \& {Kobayashi}}]{Lind2022}
{Lind}, K., {Nordlander}, T., {Wehrhahn}, A., {et~al.} 2022, \aap, 665, A33

\bibitem[{{M.~Kovalev} {et~al.}(2018){M.~Kovalev}, {S.~Brinkmann},
  {M.~Bergemann}, \& {MPIA IT-department}}]{NLTE_MPIA}
{M.~Kovalev}, {S.~Brinkmann}, {M.~Bergemann}, \& {MPIA IT-department}. 2018,
  {NLTE MPIA web server, [Online]. Available: {{http://nlte.mpia.de}} Max
  Planck Institute for Astronomy, Heidelberg.}

\bibitem[{{Matsuno} {et~al.}(2019){Matsuno}, {Aoki}, \& {Suda}}]{Matsuno2019}
{Matsuno}, T., {Aoki}, W., \& {Suda}, T. 2019, \apjl, 874, L35

\bibitem[{{Mucciarelli} {et~al.}(2021){Mucciarelli}, {Bellazzini}, \&
  {Massari}}]{Mucciarelli2021}
{Mucciarelli}, A., {Bellazzini}, M., \& {Massari}, D. 2021, \aap, 653, A90

\bibitem[{{Myeong} {et~al.}(2019){Myeong}, {Vasiliev}, {Iorio}, { Evans}, \&
  {Belokurov}}]{Myeong:2019}
{Myeong}, G.~C., {Vasiliev}, E., {Iorio}, G., { Evans}, N.~W., \& {Belokurov},
  V. 2019, \mnras, 488, 1235

\bibitem[{{Naidu} {et~al.}(2020){Naidu}, {Conroy}, {Bonaca}, {Johnson}, {Ting},
  {Caldwell}, {Zaritsky}, \& {Cargile}}]{Naidu2020}
{Naidu}, R.~P., {Conroy}, C., {Bonaca}, A., {et~al.} 2020, \apj, 901, 48

\bibitem[{{Pancino} {et~al.}(2017){Pancino}, {Romano}, {Tang},
  {Tautvai{\v{s}}ien{\.{e}}}, {Casey}, {Gruyters}, {Geisler}, {San Roman},
  {Randich}, {Alfaro}, {Bragaglia}, {Flaccomio}, {Korn}, {Recio-Blanco},
  {Smiljanic}, {Carraro}, {Bayo}, {Costado}, {Damiani}, {Jofr{\'e}}, {Lardo},
  {de Laverny}, {Monaco}, {Morbidelli}, {Sbordone}, {Sousa}, \&
  {Villanova}}]{Pancino2017}
{Pancino}, E., {Romano}, D., {Tang}, B., {et~al.} 2017, \aap, 601, A112

\bibitem[{P\'erez \& Granger(2007)}]{ipython}
P\'erez, F. \& Granger, B.~E. 2007, Computing in Science and Engineering, 9, 21

\bibitem[{{Portegies Zwart} \& {McMillan}(2000)}]{SPZ:2000}
{Portegies Zwart}, S.~F. \& {McMillan}, S. L.~W. 2000, \apjl, 528, L17

\bibitem[{{Rastello} {et~al.}(2023){Rastello}, {Iorio}, {Mapelli},
  {Arca-Sedda}, {Di Carlo}, {Escobar}, {Shenar}, \&
  {Torniamenti}}]{Rastello:2023}
{Rastello}, S., {Iorio}, G., {Mapelli}, M., {et~al.} 2023, \mnras, 526, 740

\bibitem[{{Ruiz-Lara} {et~al.}(2022){Ruiz-Lara}, {Matsuno}, {L{\"o}vdal},
  {Helmi}, {Dodd}, \& {Koppelman}}]{RL2022}
{Ruiz-Lara}, T., {Matsuno}, T., {L{\"o}vdal}, S.~S., {et~al.} 2022, \aap, 665,
  A58

\bibitem[{{Sbordone} {et~al.}(2004){Sbordone}, {Bonifacio}, {Castelli}, \&
  {Kurucz}}]{sbordone04}
{Sbordone}, L., {Bonifacio}, P., {Castelli}, F., \& {Kurucz}, R.~L. 2004,
  Memorie della Societa Astronomica Italiana Supplementi, 5, 93

\bibitem[{{Schlegel} {et~al.}(1998){Schlegel}, {Finkbeiner}, \&
  {Davis}}]{Schlegel1998}
{Schlegel}, D.~J., {Finkbeiner}, D.~P., \& {Davis}, M. 1998, \apj, 500, 525

\bibitem[{{Simon}(2019)}]{Simon:2019}
{Simon}, J.~D. 2019, \araa, 57, 375

\bibitem[{{Sneden}(1973)}]{Sneden1973}
{Sneden}, C.~A. 1973, PhD thesis, University of Texas, Austin

\bibitem[{{Tanikawa} {et~al.}(2024){Tanikawa}, {Cary}, {Shikauchi}, {Wang}, \&
  {Fujii}}]{Tanikawa2024}
{Tanikawa}, A., {Cary}, S., {Shikauchi}, M., {Wang}, L., \& {Fujii}, M.~S.
  2024, \mnras, 527, 4031

\bibitem[{{Vasiliev} \& {Baumgardt}(2021)}]{Vasiliev21}
{Vasiliev}, E. \& {Baumgardt}, H. 2021, \mnras, 505, 5978

\bibitem[{{Vernet} {et~al.}(2011){Vernet}, {Dekker}, {D'Odorico}, {Kaper},
  {Kjaergaard}, {Hammer}, {Randich}, {Zerbi}, {Groot}, {Hjorth}, {Guinouard},
  {Navarro}, {Adolfse}, {Albers}, {Amans}, {Andersen}, {Andersen}, {Binetruy},
  {Bristow}, {Castillo}, {Chemla}, {Christensen}, {Conconi}, {Conzelmann},
  {Dam}, {de Caprio}, {de Ugarte Postigo}, {Delabre}, {di Marcantonio},
  {Downing}, {Elswijk}, {Finger}, {Fischer}, {Flores}, {Fran{\c{c}}ois},
  {Goldoni}, {Guglielmi}, {Haigron}, {Hanenburg}, {Hendriks}, {Horrobin},
  {Horville}, {Jessen}, {Kerber}, {Kern}, {Kiekebusch}, {Kleszcz}, {Klougart},
  {Kragt}, {Larsen}, {Lizon}, {Lucuix}, {Mainieri}, {Manuputy}, {Martayan},
  {Mason}, {Mazzoleni}, {Michaelsen}, {Modigliani}, {Moehler}, {M{\o}ller},
  {Norup S{\o}rensen}, {N{\o}rregaard}, {P{\'e}roux}, {Patat}, {Pena}, {Pragt},
  {Reinero}, {Rigal}, {Riva}, {Roelfsema}, {Royer}, {Sacco}, {Santin},
  {Schoenmaker}, {Spano}, {Sweers}, {Ter Horst}, {Tintori}, {Tromp}, {van
  Dael}, {van der Vliet}, {Venema}, {Vidali}, {Vinther}, {Vola}, {Winters},
  {Wistisen}, {Wulterkens}, \& {Zacchei}}]{Vernet:2011}
{Vernet}, J., {Dekker}, H., {D'Odorico}, S., {et~al.} 2011, \aap, 536, A105

\bibitem[{Walt {et~al.}(2011)Walt, Colbert, \& Varoquaux}]{numpy}
Walt, S. v.~d., Colbert, S.~C., \& Varoquaux, G. 2011, Computing in Science and
  Engg., 13, 22

\bibitem[{{Ying} {et~al.}(2023){Ying}, {Chaboyer}, {Boudreaux}, {Slaughter},
  {Boylan-Kolchin}, \& {Weisz}}]{Ying:2023}
{Ying}, J.~M., {Chaboyer}, B., {Boudreaux}, E.~M., {et~al.} 2023, \aj, 166, 18

\end{thebibliography}

\appendix
\section{Derivation of stellar parameters and chemical abundances of ED-2 stars}
\label{sec:appendix}

We obtained spectra for 3 stars as part of the follow-up of the ED-2 stream in
period 111 (April -- September 2023; proposal submitted in September 2022;
program 0111.D-0263(A), PI:Dodd) with the optical spectrograph, UVES mounted at
the Very Large Telescope (VLT) of the European Southern Observatory. The
observations were performed with UVES in Dichroic mode adopting the standard
settings Dic 1 Blue Arm CD2 390 (326–454 nm) and Dic 1 Red Arm CD3 580 (476--684
nm) and with the 0.7'' slit width, thus yielding a resolution of $R\sim 55 000$,
and S/N $\ge$ 15 for the Blue Arm and S/N$\ge$30 for the Red Arm on average. We
have also used ESO archival data from the programs 0109.B-0522(A) and
167.D-0173(A), for another ED-2 member, {\tt source\_ID} 4479226310758314496.
Additionally, we observed 7 ED-2 stars with X-Shooter at the VLT in period 112
(October 2023 -- March 2024; proposal submitted in March 2023; program
112.25ZW.001; PI: Balbinot). In all cases, we used the phase 3 data products
provided by ESO for further analysis. 

 For the UVES spectra we derived the chemical abundances of the stars using the
 1D LTE spectral synthesis code MOOG \citep{Sneden1973} with the grid of MARCS
 model atmospheres \citep{Gustafsson2008}. Stellar parameters ($T_{\rm eff}$ and
 $\log g$) were determined from dereddened photometry and astrometry; $T_{\rm
 eff}$ was obtained from the $G-K_s$ color using the relation from
 \citet{Mucciarelli2021}, and $\log g$ was obtained from the $K_s$ magnitude
 together with the bolometric correction of \citet{Casagrande2014} and an
 assumption of a mass of $0.8\,\mathrm{M}_\odot$. The extinction was taken from
 \citet{Green2019} where available and \citet{Schlegel1998} otherwise. We
 measured abundances of Mg and Fe through equivalent widths analysis and of Na, Al, Mn, 
 Ba, and Eu through spectral synthesis with hyperfine structure splitting
 included, and applied non-LTE corrections of \citet{Lind2022} to the Na and Al
 abundances. We simply averaged the line-by-line abundances to obtain the final
 abundance of each element. We estimate the uncertainties from the sample
 standard deviation of the line-by-line abundances ($\sigma$) and the number of
 lines ($N$) as $\sigma/\sqrt{N}$ when $N>3$; otherwise, we replace the $\sigma$
 with that of neutral iron. We additionally consider the uncertainties due to
 the stellar parameters. We report the measured abundances in Table~\ref{table:1}.

For the X-Shooter spectra, we initially stack individual radial velocity
corrected exposures. We assume atmospheric parameters from \citet{Andrae:2023},
with the exception of star 3757312745743087232, where Gaia XP spectra was used
instead. We synthesised H${\alpha}$ and H${\beta}$ non-local thermodynamic
equilibrium (NLTE) line profiles using the tools provided by \cite{NLTE_MPIA}.
We find a that the adopted values for T$_{\rm eff}$ and $\log$(g) adequately
reproduce the wings of the Balmer lines. The spectra for each star was
normalized assuming a \feh = -2.5 template in the range between 330nm to 1100nm.
Finally, while keeping the atmospheric parameters constant we derived Fe
abundances using the {\tt SYNTHE} transfer code \citep{sbordone04}, assuming the
\texttt{ATLAS 9} models \citep{kurucz05}, and atomic data from \citet{vald}. We
do so by minimising the $\chi^2$ between observed and synthetic fluxes around a
set of selected Fe features. We report the Fe abundances and their associated
uncertainty in Table \ref{tab:specXS}.

\begin{table*}
\centering
\caption{Stellar parameters and chemical abundances for ED-2 stars. The 3 ED-2
stars in our UVES program, supplemented by an ED-2 star from the ESO archive
(second entry) are shown in the first part. The bottom two entries correspond to
the binary star of \gaia BH3, as estimated following the procedure described in
this paper and as presented in \citet{Panuzzo:2024}, respectively. 
}
\scalebox{0.72}{
\begin{tabular}{lccccccccccc}  
\hline
 \texttt{source\_id}& $T_{\rm eff}$ & $\log g$ & $v_t$                    & $[\mathrm{Fe/H}]_{\rm I}$ & $[\mathrm{Fe/H}]_{\rm II}$ & $[\mathrm{Na/Fe}]$ & $[\mathrm{Mg/Fe}]$ & $[\mathrm{Al/Fe}]$ &  $[\mathrm{Mn/Fe}]$ & $[\mathrm{Ba/Fe}]$ & $[\mathrm{Eu/Fe}]$ \\ 
-- &   [K]           &   --       &  $\mathrm{[km\,s^{-1}]}$ &        --                 &        --                  &      --            &         --         &        --          &    --              &        --          \\ 
\hline                     
\hline

4245522468554091904                                 & 6657 & 4.30 & 1.68 & -2.48 $\pm$ 0.05 & -2.46 $\pm$ 0.10 & -0.25 $\pm$ 0.10 & 0.18 $\pm$ 0.07 & -0.22 $\pm$ 0.07 & -0.45 $\pm$ 0.12 & -0.13 $\pm$ 0.15 &                 \\
4479226310758314496                                 & 5974 & 4.53 & 1.12 & -2.70 $\pm$ 0.05 & -2.57 $\pm$ 0.06 & -0.23 $\pm$ 0.07 & 0.30 $\pm$ 0.08 & -0.34 $\pm$ 0.07 & -0.52 $\pm$ 0.09 & -0.35 $\pm$ 0.10 &                 \\
6632335060231088896                                 & 5620 & 3.60 & 1.32 & -2.54 $\pm$ 0.05 & -2.44 $\pm$ 0.04 & -0.07 $\pm$ 0.08 & 0.33 $\pm$ 0.06 & -0.32 $\pm$ 0.07 & -0.41 $\pm$ 0.06 & -0.17 $\pm$ 0.08 & 0.61 $\pm$ 0.08 \\
6746114585056265600                                 & 6110 & 4.55 & 1.28 & -2.57 $\pm$ 0.05 & -2.44 $\pm$ 0.02 & -0.25 $\pm$ 0.08 & 0.29 $\pm$ 0.06 & -0.27 $\pm$ 0.08 & -0.43 $\pm$ 0.06 & -0.24 $\pm$ 0.09 &                 \\ \hline
\textbf{4318465066420528000} \tablefootmark{a}      & 5445 & 3.04 & 1.60 & -2.33 $\pm$ 0.05 & -2.35 $\pm$ 0.05 & -0.15 $\pm$ 0.09 & 0.38 $\pm$ 0.04 & -0.17 $\pm$ 0.09 & -0.14 $\pm$ 0.10 & -0.04 $\pm$ 0.10 & 0.58 $\pm$ 0.10 \\
\textbf{4318465066420528000$^{*}$}\tablefootmark{a} & 5211 & 2.93 & 1.19 & -2.56 $\pm$ 0.12 & -2.49 $\pm$ 0.12 &                  & 0.48 $\pm$ 0.05 &                  &                  &  0.11 $\pm$ 0.13 & 0.52 $\pm$ 0.05 \\\hline
\end{tabular}
}
\tablefoot{\tablefoottext{a}{We note that the difference $T_{\rm eff}$ between
the two studies is due to different extinction maps used and is responsible for
the relatively large difference in $[\mathrm{Fe/H}]_{\rm I}$. The differences in
the other ratios are likely due to systematic errors that are not accounted for
in the quoted uncertainties, as the abundance ratios are rather insensitive to
the choice of $T_{\rm eff}$. Sources for such systematic uncertainties include
the use of different linelists, radiative transfer codes, and  atmospheric
models.
}}
\label{table:1}
\end{table*}

\begin{table*}
\caption{Stellar parameters and [Fe/H] for the X-Shooter sample. The reported
$T_{\rm eff}$ and $\log g$ are from \citet{Andrae:2023}, except for star
\texttt{source\_id} 3757312745743087232 where the atmospheric parameters were
infered from the Gaia XP spectra. We add a $*$ to the {\tt source\_id} to mark
stars not original in \citet{Dodd:2022}.}
\label{tab:specXS}
\centering
\scalebox{0.8}{
\begin{tabular}{lccc}
\hline
 \texttt{source\_id} & $T_{\rm eff}$ & $\log g$ & $[\mathrm{Fe/H}]_{\mathrm{I}}$   \\
     --              &    [K]        &   --     &          --           \\
\hline
\hline
  3549718318990080896   & 5319 & 3.3 & -2.52 $\pm$ 0.08  \\
  3869876996687740032*  & 5372 & 3.4 & -2.50 $\pm$ 0.10  \\
  6747065215934660608   & 5535 & 4.3 & -2.53 $\pm$ 0.08  \\
  6646097819069706624   & 5653 & 4.4 & -2.54 $\pm$ 0.10  \\
  3473979147705211776   & 5733 & 4.4 & -2.57 $\pm$ 0.14  \\
  3757312745743087232   & 6591 & 4.2 & -2.23 $\pm$ 0.41  \\
\hline
    \end{tabular}
}
\end{table*}


\end{document}